\documentclass[conference]{IEEEtran}
\IEEEoverridecommandlockouts
\usepackage{cite}
\usepackage{amsmath,amssymb,amsfonts}
\usepackage{algorithm} 
\usepackage{algpseudocode}
\usepackage{graphicx}
\usepackage{textcomp}
\usepackage{xcolor}
\def\BibTeX{{\rm B\kern-.05em{\sc i\kern-.025em b}\kern-.08em
    T\kern-.1667em\lower.7ex\hbox{E}\kern-.125emX}}

\newcounter{MYtempeqncnt}

\allowdisplaybreaks[4]

\begin{document}

\title{Joint Beamforming for Continuous Transmissive RIS-Enabled Multiuser Communications}
\author{\IEEEauthorblockN{Beining Han, Shumeng Zhang, Deyou Zhang, Qingchao Li, Jun Liu, and Chuang Shi}
\IEEEauthorblockA{\textit{School of Electronic and Information Engineering, Beihang University, Beijing 100191, China} \\
email: \{beininghan, simoncaffrey, deyou, qingchaoli, liujun2019, shichuang\}@buaa.edu.cn}
}

\maketitle

\begin{abstract}
This paper studies continuous transmissive reconfigurable intelligent surface (CT-RIS)-enabled multiuser downlink communications, where passive beamforming is characterized by a phase function defined over the continuous aperture. We employ a distance-dependent spherical-wave model for the base station (BS)-to-CT-RIS link and introduce a finite-path model based on field-response vectors for the CT-RIS-to-user links. The resulting weighted sum rate maximization problem jointly optimizes the finite-dimensional BS precoders and the infinite-dimensional CT-RIS phase function, giving rise to a mixed finite- and infinite-dimensional non-convex optimization problem. To address this problem, we develop an alternating optimization algorithm that combines a closed-form BS precoder update with a functional phase update derived using the calculus of variations and minorization-maximization. Numerical results demonstrate that the proposed CT-RIS consistently outperforms the considered benchmarks, with the performance gains attributed to spatially continuous phase control and a larger effective receiving area.
\end{abstract}


\section{Introduction}
Reconfigurable intelligent surfaces (RISs) enable programmable control of wireless propagation environments \cite{9424177}. Conventional transmissive RISs (T-RISs) typically adopt an element-based architecture comprising a finite number of individually controlled elements arranged across a panel \cite{10680462}. By jointly designing the element-wise T-RIS phase shifts and base station (BS) digital precoding, T-RIS-enabled systems can improve communication performance without requiring active radio-frequency chains at the T-RIS \cite{10177872}.

Existing studies on such element-based T-RISs have investigated hardware implementations \cite{10680462}, beamforming optimization \cite{10740042}, multiuser communications \cite{11342402}, and simultaneous wireless information and power transfer \cite{9983541}. In these studies, passive beamforming is typically formulated as a finite-dimensional optimization problem over an element-wise phase-shift vector. While this formulation is consistent with conventional element-based architectures, it does not directly represent the aperture response as a function of a continuous spatial coordinate. Recent studies on holographic multiple-input multiple-output and continuous-aperture arrays (CAPAs) have developed functional-domain models in which aperture responses are described as functions of continuous spatial coordinates \cite{9136592,9443506,10910020}, motivating a continuous-aperture formulation of T-RISs. However, unlike CAPAs that directly transmit or receive signals, a continuous T-RIS (CT-RIS) passively modulates the electromagnetic field incident from an external transmitter. Its modeling therefore requires characterizing the propagation on both sides of the aperture and coherently integrating their pointwise contributions. Channel modeling and joint beamforming optimization for CT-RIS-enabled systems remain largely unexplored.

To address this gap, this paper investigates CT-RIS-enabled multiuser downlink communications, where the CT-RIS phase response is represented by a function defined over the continuous aperture. We employ a distance-dependent spherical-wave model for the BS-to-CT-RIS link and introduce a finite-path model based on field-response vectors (FRVs) for the CT-RIS-to-user links \cite{10318061}. The pointwise responses of the two propagation links are combined with the CT-RIS phase response and coherently integrated over the aperture to obtain the equivalent BS-to-user channels. Based on the resulting channel model, we formulate a weighted sum rate (WSR) maximization problem that jointly optimizes the finite-dimensional BS precoders and the infinite-dimensional CT-RIS phase function, giving rise to a mixed finite- and infinite-dimensional non-convex optimization problem.

To solve this problem, we develop an alternating optimization (AO) framework. By applying the Lagrangian dual transform and the complex quadratic transform, the original WSR objective is reformulated into a blockwise tractable form. The auxiliary variables are updated in closed form, while the BS precoders admit a closed-form expression derived from the Karush--Kuhn--Tucker (KKT) conditions, with the associated scalar dual variable determined via a one-dimensional bisection search. For the infinite-dimensional CT-RIS phase function, the calculus of variations (CoV) is employed to characterize its functional stationarity condition, while a minorization-maximization (MM) method constructs a tight minorizing surrogate that yields a pointwise phase update with guaranteed monotonic improvement of the transformed objective. In the numerical evaluation, a subaperture CT-RIS is introduced as an intermediate benchmark to decompose the performance gain of the proposed CT-RIS over the conventional element-based T-RIS into contributions from spatially continuous phase control and a larger effective receiving area.

\textit{Notation}: Scalars, vectors, matrices, and sets are denoted by italic, bold lowercase, bold uppercase, and calligraphic letters, respectively. The symbols $\mathbb{R}$ and $\mathbb{C}$ denote the sets of real and complex numbers, respectively, $j = \sqrt{-1}$ is the imaginary unit, and $\mathbf I_N$ denotes the $N \times N$ identity matrix. The operators $(\cdot)^T$, $(\cdot)^*$, $(\cdot)^H$, $\operatorname{Re}\{\cdot\}$, $\operatorname{Im}\{\cdot\}$, $\vert{}\cdot\vert{}$, $\Vert{}\cdot\Vert{}$, $\operatorname{Tr}(\cdot)$, and $\mathbb{E}\{\cdot\}$ represent transpose, complex conjugate, conjugate transpose, real part, imaginary part, absolute value (or the measure of a set), Euclidean norm, trace, and expectation, respectively. $\langle u,v\rangle = \int_{\mathcal S} u^*(\mathbf s) v(\mathbf s) \mathrm{d} \mathbf s$, $u, v \in L^2(\mathcal{S})$.

\section{System Model and Problem Formulation}
\begin{figure}[t]
	\centering
    \includegraphics[width=0.85\columnwidth]{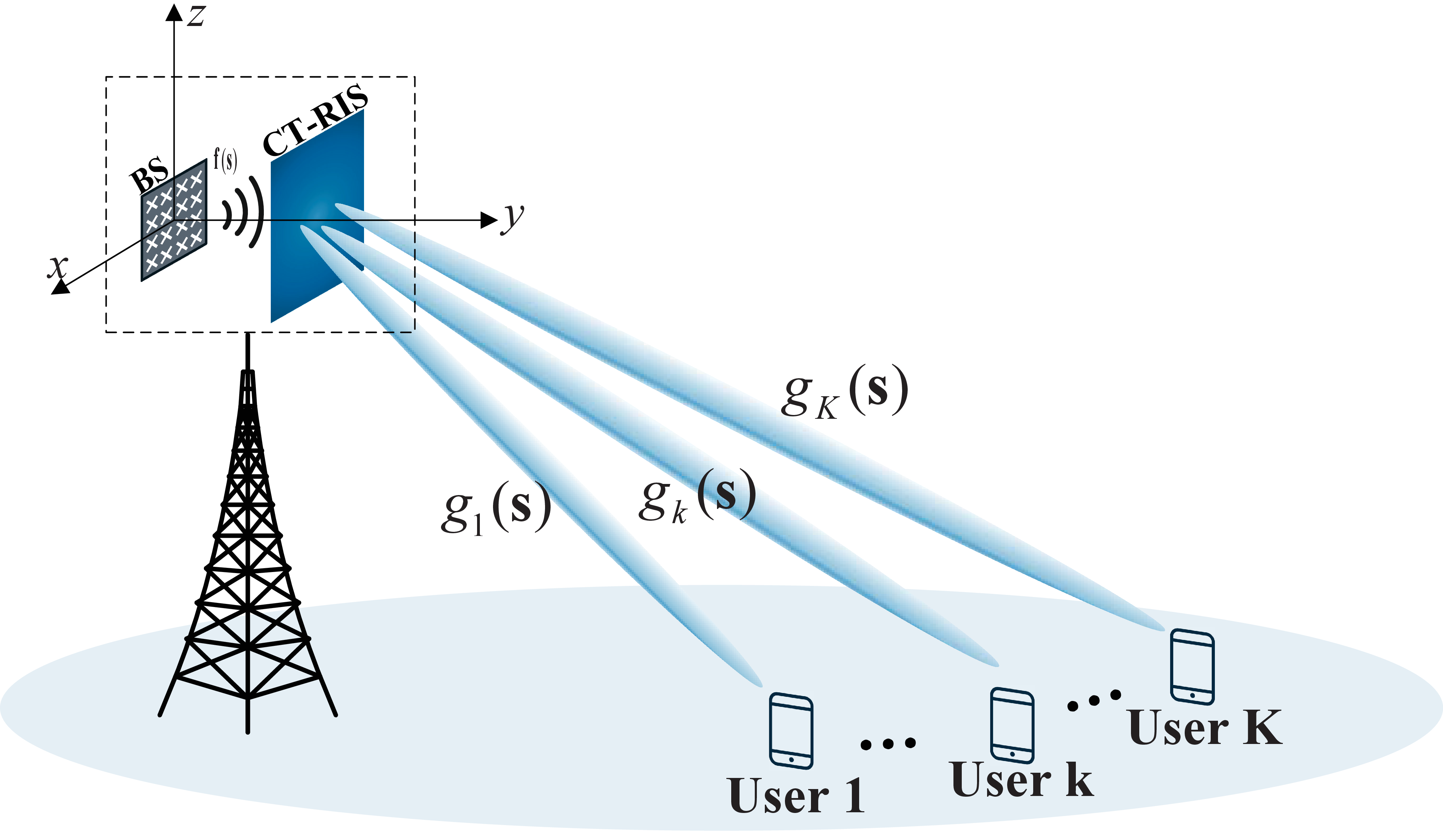}
	\caption{Illustration of the CT-RIS-enabled multiuser downlink system.}
	\label{fig:system}
    \vskip-10pt
\end{figure}

As illustrated in Fig.~\ref{fig:system}, we consider a CT-RIS-enabled multiuser downlink system in which an $N$-antenna uniform planar array (UPA) at the BS serves $K$ single-antenna users through a CT-RIS. The direct BS-to-user links are assumed to be blocked or negligible.

\subsection{Transmit--Receive Signal Model}

The BS employs linear digital precoding to generate the transmitted signal vector
\begin{equation}
	\mathbf{x} = \mathbf W \mathbf u \in \mathbb C^{N \times 1},
\end{equation}
where $\mathbf u=[u_1,\ldots,u_K]^T\in\mathbb C^{K\times1}$ denotes the symbol vector satisfying
$\mathbb E\{\mathbf u\mathbf u^H\}=\mathbf I_K$, and
$\mathbf W=[\mathbf w_1,\ldots,\mathbf w_K]\in\mathbb C^{N\times K}$ is the digital precoding matrix, with $\mathbf w_k$ denoting the precoding vector for user $k$.

Let $\mathcal S\subset\mathbb R^3$ denote the CT-RIS aperture and $\mathbf s\in\mathcal S$ an arbitrary aperture point. We denote by
$\mathbf f(\mathbf s)\in\mathbb C^{N\times1}$ the channel vector from the BS antennas to aperture point $\mathbf s$, and by
$g_k(\mathbf s)\in\mathbb C$ the channel response from $\mathbf s$ to user $k$. The specific models of $\mathbf f(\mathbf s)$ and $g_k(\mathbf s)$ are presented in the next subsection.

The incident complex baseband signal at aperture point $\mathbf s$ is
\begin{equation}
	y_{\mathrm T}(\mathbf s) = \mathbf f^T(\mathbf s)\mathbf x.
\end{equation}
The CT-RIS phase response is represented by a function
$\theta(\mathbf s)\in[-\pi,\pi)$ defined over the continuous aperture. Under an ideal lossless phase-only transmission model, the transmission coefficient at $\mathbf s$ is $e^{j\theta(\mathbf s)}$. Accordingly, the complex baseband signal immediately after transmission through the CT-RIS is
\begin{equation}
	x_{\mathrm T}(\mathbf s) = y_{\mathrm T}(\mathbf s)e^{j\theta(\mathbf s)}.
\end{equation}

By coherently integrating the pointwise contributions over the CT-RIS aperture, the received signal at user $k$ is
\begin{align}
	y_k & = \eta \int_{\mathcal S} g_k(\mathbf s) x_{\mathrm T}(\mathbf s) \,\mathrm d\mathbf s + n_k \nonumber \\[2ex]
	& = \eta \int_{\mathcal S} g_k(\mathbf s) \mathbf f^T(\mathbf s) \mathbf x e^{j\theta(\mathbf s)} \,\mathrm d\mathbf s + n_k,
	\label{eq:received_signal_integral}
\end{align}
where $n_k \sim \mathcal{CN}(0,\sigma_k^2)$ denotes the additive white Gaussian noise, and $\eta = \frac{1}{A_{\mathrm{ref}}} = \frac{4 \pi}{\lambda_c^2}$ is the surface-response normalization coefficient, with $\lambda_c$ denoting the carrier wavelength and $A_{\mathrm{ref}} = \lambda_c^2 / (4\pi)$ denoting the effective aperture of an ideal unit-gain isotropic point element. Since $\mathbf f(\mathbf s)$ and $g_k(\mathbf s)$ are dimensionless channel responses while ${\mathrm d}{\mathbf s}$ has units of area, multiplication by $\eta$ removes the area dimension introduced by the surface integration and yields a dimensionless equivalent channel. The same normalization is adopted for the element-based T-RIS benchmark.

Define the equivalent BS-to-user $k$ channel as
\begin{align}
	\mathbf h_k^H
	\triangleq
	\eta
	\int_{\mathcal S}
	g_k(\mathbf s)\mathbf f^T(\mathbf s)
	e^{j\theta(\mathbf s)} \,\mathrm d\mathbf s.
	\label{eq:equivalent_channel}
\end{align}
The received signal can then be compactly expressed as
\begin{align}
	y_k
	=
	\mathbf h_k^H\mathbf w_k u_k
	+
	\sum_{i\ne k}
	\mathbf h_k^H\mathbf w_i u_i
	+
	n_k.
	\label{eq:yk}
\end{align}

\subsection{Channel Modeling}
\subsubsection{BS-to-CT-RIS Link}
Let $\mathbf p_n\in\mathbb R^3$ denote the position of the $n$-th BS antenna element. To account for wavefront curvature across the CT-RIS aperture, we retain the exact distance between each BS antenna element and each aperture point. Under a scalar free-space line-of-sight (LoS) propagation model, the complex baseband channel response from the $n$-th BS antenna element to $\mathbf s$ is
\begin{align}
	f_n(\mathbf s)
	=
	\frac{\lambda_c}{4\pi d_n(\mathbf s)}
	e^{-j\frac{2\pi}{\lambda_c}d_n(\mathbf s)},
	\quad n=1,\ldots,N,
	\label{eq:BS_CT_channel}
\end{align}
where $d_n(\mathbf s) = \|\mathbf s-\mathbf p_n\|_2$ is the corresponding propagation distance. This model retains the point-dependent amplitude and phase variations across the aperture without invoking a plane-wave approximation. The BS-to-aperture channel vector is therefore $\mathbf f(\mathbf s)\triangleq[f_1(\mathbf s),\ldots,f_N(\mathbf s)]^T\in\mathbb C^{N\times1}$.

\begin{figure*}[t]
\normalsize
\setcounter{MYtempeqncnt}{\value{equation}}
\setcounter{equation}{13}
\begin{equation}\label{eq:transformed_function}
F_0
\left(
\mathbf W,\theta,
\boldsymbol{\mu},
\boldsymbol{\lambda}
\right)
\triangleq
\frac{1}{\ln 2}
\sum_{k=1}^{K}
\alpha_k
\left[
\ln(\mu_k^2)-\mu_k^2+1
+2\mu_k\operatorname{Re}
\left\{
\lambda_k^*\mathbf h_k^H\mathbf w_k
\right\}
-|\lambda_k|^2
\left(
\sum_{i=1}^{K}
|\mathbf h_k^H\mathbf w_i|^2
+\sigma_k^2
\right)
\right].
\end{equation}
\hrulefill
\vskip-12pt
\end{figure*}
\setcounter{equation}{\value{MYtempeqncnt}}

\subsubsection{FRV-Based CT-RIS-to-User Links}

Following the FRV framework in \cite{10318061}, we model each CT-RIS-to-user link using a finite number of propagation paths. The far-field plane-wave approximation is assumed to hold over the CT-RIS aperture. Let
$\mathbf s_0=[x_0,y_0,z_0]^T$ denote the aperture center, and let $L_k$ denote the number of propagation paths between the CT-RIS and user $k$.

For the $l$-th path, define the unit propagation-direction vector toward the CT-RIS as
\begin{align}
	\mathbf q_{k,l}
	=
	\begin{bmatrix}
		\cos\vartheta_{k,l}\cos\phi_{k,l}\\
		\cos\vartheta_{k,l}\sin\phi_{k,l}\\
		\sin\vartheta_{k,l}
	\end{bmatrix},
\end{align}
where $\vartheta_{k,l}$ and $\phi_{k,l}$ denote the elevation and azimuth angles, respectively. The path-length offset at aperture point $\mathbf s$ relative to the aperture center is $\rho_{k,l}(\mathbf s) = (\mathbf s-\mathbf s_0)^T\mathbf q_{k,l}$. Accordingly, the FRV associated with the CT-RIS-to-user $k$ link is defined as 
\begin{equation}
    \mathbf a_k(\mathbf s)=[e^{j\frac{2\pi}{\lambda_c}\rho_{k,1}(\mathbf s)},\ldots,e^{j\frac{2\pi}{\lambda_c}\rho_{k,L_k}(\mathbf s)}]^T\in\mathbb C^{L_k\times1}.
\end{equation}

Let $\boldsymbol{\beta}_k=[\beta_{k,1}, \ldots, \beta_{k,L_k}]^T \in\mathbb C^{L_k\times1}$ collect the complex path gains, including the path attenuation and reference phase at the aperture center. The channel response from aperture point $\mathbf s$ to user $k$ is then expressed as
\begin{align}
	g_k(\mathbf s) =
	\mathbf a_k^H(\mathbf s)\boldsymbol{\beta}_k,
	\label{eq:FRV_channel}
\end{align}
which can accommodate both LoS and scattered propagation paths.

\subsection{Problem Formulation}
We assume that the BS has access to the geometric and path parameters required to construct $\mathbf f(\mathbf s)$ and $\{g_k(\mathbf s)\}_{k=1}^{K}$, such that the equivalent channels can be evaluated for any feasible CT-RIS phase function. Based on the received signal in \eqref{eq:yk}, the signal-to-interference-plus-noise ratio (SINR) at user $k$ for decoding $u_k$ is given by
\begin{align}
	\gamma_k
	=
	\frac{
		\left|\mathbf h_k^H\mathbf w_k\right|^2
	}{
		\sum_{i=1, i\neq k}^{K}
		\left|\mathbf h_k^H\mathbf w_i\right|^2
		+
		\sigma_k^2
	}.
	\label{eq:SINR}
\end{align}
Accordingly, the achievable rate of user $k$ is $R_k=\log_2(1+\gamma_k)$ in bit/s/Hz.

We jointly design the BS precoding matrix $\mathbf W$ and the CT-RIS phase function $\theta(\cdot)$ to maximize the WSR. The corresponding optimization problem is formulated as
\begin{subequations}
	\label{eq:WSR_problem}
	\begin{align}
		\max_{\mathbf W,\,\theta(\cdot)}
		\quad
		&
		\sum_{k=1}^{K}
		\alpha_k\log_2(1+\gamma_k)
		\label{eq:WSR_objective}
		\\
		\mathrm{s.t.}\quad
		&
		\sum_{k=1}^{K}
		\|\mathbf w_k\|_2^2
		\leq
		P_{\mathrm T},
		\label{eq:power_constraint}
		\\[1ex]
		&
		\theta(\mathbf s)\in[-\pi,\pi),
		\quad
		\forall\,\mathbf s\in\mathcal S,
		\label{eq:phase_constraint}
	\end{align}
\end{subequations}
where $\alpha_k>0$ denotes the priority weight assigned to user $k$, and $P_{\mathrm T}$ is the maximum transmit power of the BS.

Problem~\eqref{eq:WSR_problem} jointly optimizes the finite-dimensional BS precoding matrix and the infinite-dimensional phase function defined over the CT-RIS aperture. The problem is non-convex because the user SINRs are coupled through both variable blocks and the associated transmission coefficient $e^{j\theta(\mathbf s)}$ is subject to a pointwise unit-modulus constraint. To address these challenges, the following section develops an alternating optimization framework for iteratively updating the BS precoders and the CT-RIS phase function.

\section{AO-CoV-MM Algorithm for Joint Precoding and Functional Phase Optimization}

To solve Problem~\eqref{eq:WSR_problem}, we develop an AO-CoV-MM algorithm that iteratively performs three updates: the auxiliary-variable update, the BS precoder update, and the CT-RIS phase-function update. The auxiliary variables are introduced to obtain an equivalent and blockwise tractable reformulation of the WSR objective, while the CT-RIS phase function is updated using CoV and MM.

\subsection{Equivalent Reformulation of the WSR Problem}

The WSR objective contains logarithmic functions of fractional SINR terms, which couple the BS precoders and the CT-RIS phase function. By applying the Lagrangian dual transform and the complex quadratic transform \cite{8314727,8310563}, Problem~\eqref{eq:WSR_problem} can be equivalently reformulated as
\begin{subequations}
	\label{eq:transformed_problem}
	\begin{align}
		\max_{\mathbf W,\,\theta(\cdot),
		\,\boldsymbol{\mu},\,\boldsymbol{\lambda}}
		\quad
		&
		F_0
		\left(
		\mathbf W,\theta,
		\boldsymbol{\mu},
		\boldsymbol{\lambda}
		\right)
		\label{eq:transformed_objective}
		\\
		\mathrm{s.t.}\quad
		&
		\sum_{k=1}^{K}
		\|\mathbf w_k\|_2^2
		\leq
		P_{\mathrm T},
		\label{eq:transformed_power_constraint}
		\\[1ex]
		&
		\theta(\mathbf s)\in[-\pi,\pi),
		\quad
		\forall\,\mathbf s\in\mathcal S,
		\label{eq:transformed_phase_constraint}
		\\[1ex]
		&
		\mu_k\geq1,\quad
		\lambda_k\in\mathbb C,
		\quad
		k=1,\ldots,K.
		\label{eq:transformed_auxiliary_constraints}
	\end{align}
\end{subequations}
Here, $\boldsymbol{\mu}=[\mu_1,\ldots,\mu_K]^T$ and
$\boldsymbol{\lambda}=[\lambda_1,\ldots,\lambda_K]^T$
collect the auxiliary variables introduced by the Lagrangian dual transform and the complex quadratic transform. $F_0(\mathbf W,\theta,\boldsymbol{\mu},\boldsymbol{\lambda})$
is defined in \eqref{eq:transformed_function}, where the dependence of $\mathbf h_k$ on $\theta(\cdot)$ is suppressed for notational simplicity. For any fixed feasible $\mathbf W$ and $\theta(\cdot)$, maximizing
$F_0$ over $\boldsymbol{\mu}$ and $\boldsymbol{\lambda}$ recovers the original WSR objective. Hence, \eqref{eq:WSR_problem} and \eqref{eq:transformed_problem} are equivalent.

\subsection{Block Update Rules}
At each AO iteration, the auxiliary variables, the BS precoders, and the CT-RIS phase function are updated sequentially. The auxiliary-variable and BS-precoding subproblems are optimally solved for fixed values of the remaining variables, whereas the CT-RIS phase function is updated through an MM step that guarantees a nondecreasing transformed objective.

\subsubsection{Auxiliary Variable Update}
For fixed $\mathbf W$ and $\theta(\cdot)$, the optimal auxiliary variables obtained from the Lagrangian dual and complex quadratic transforms are given by \cite{8314727,8310563}
\setcounter{equation}{14}
\begin{subequations}
	\label{eq:auxiliary_updates}
	\begin{align}
		\mu_k^\star
		&=
		\sqrt{1+\gamma_k},
		\label{eq:mu_update}
		\\[1ex]
		\lambda_k^\star
		&=
		\frac{
		\mu_k^\star\mathbf h_k^H\mathbf w_k
		}{
		\sum_{i=1}^{K}
		\left|
		\mathbf h_k^H\mathbf w_i
		\right|^2
		+
		\sigma_k^2
		},
		\label{eq:lambda_update}
	\end{align}
\end{subequations}
where $\gamma_k$ given by \eqref{eq:SINR} is evaluated using the current BS precoders and CT-RIS phase function.

\subsubsection{BS Precoder Update}
For fixed $\theta(\cdot)$ and the updated auxiliary variables $\{\mu_k,\lambda_k\}_{k=1}^{K}$, the terms in
$F_0$ that depend on $\mathbf W$ yield the following subproblem:
\begin{subequations}
	\label{eq:precoder_subproblem}
	\begin{align}
		\max_{\mathbf W}
		\quad
		&
		\sum_{k=1}^{K}
		F_k(\mathbf w_k)
		\label{eq:precoder_objective}
		\\
		\mathrm{s.t.}\quad
		&
		\sum_{k=1}^{K}
		\|\mathbf w_k\|_2^2
		\leq
		P_{\mathrm T},
		\label{eq:precoder_power_constraint}
	\end{align}
\end{subequations}
where $F_k(\mathbf w_k)\triangleq2\operatorname{Re}\left\{A_k\mathbf h_k^H\mathbf w_k\right\}-\mathbf w_k^H\mathbf Q\mathbf w_k$,
with $A_k\triangleq\frac{\alpha_k\mu_k\lambda_k^*}{\ln 2}$, $B_k\triangleq\frac{\alpha_k|\lambda_k|^2}{\ln 2}$, and $\mathbf Q\triangleq\sum_{i=1}^{K}B_i\mathbf h_i\mathbf h_i^H$.
Since $\mathbf Q\succeq\mathbf 0$, problem~\eqref{eq:precoder_subproblem} is a concave quadratic maximization problem over a convex feasible set.

Introducing the dual variable $\nu \geq 0$ associated with the transmit-power constraint, the Lagrangian is
\begin{equation}
	\mathcal L(\mathbf W,\nu)
	=
	\sum_{k=1}^{K}
	F_k(\mathbf w_k)
	-
	\nu
	\left(
	\sum_{k=1}^{K}
	\|\mathbf w_k\|_2^2
	-
	P_{\mathrm T}
	\right).
	\label{eq:precoder_lagrangian}
\end{equation}
The corresponding KKT conditions include $(\mathbf Q+\nu\mathbf I_N)\mathbf w_k= A_k^* \mathbf h_k$ and $\nu \left(\sum_{k=1}^{K} \|\mathbf w_k\|_2^2 - P_{\mathrm T} \right) = 0$. For a given $\nu \geq 0$, the stationarity condition yields\footnote{When $\nu = 0$ and $\mathbf Q$ is singular, the matrix inverse in \eqref{eq:w_update} is interpreted as the Moore--Penrose pseudoinverse.}
\begin{align}
	\mathbf w_k(\nu)
	=
	A_k^*
	\left(
	\mathbf Q+\nu\mathbf I_N
	\right)^{-1}
	\mathbf h_k,
	\quad
	\forall k.
	\label{eq:w_update}
\end{align}
The optimal dual variable is determined as follows. Define $P(\nu)\triangleq\sum_{k=1}^{K}\|\mathbf w_k(\nu)\|_2^2$. If $P(0)\leq P_{\mathrm T}$, then $\nu^\star=0$; otherwise, $\nu^\star>0$ is the unique solution to $P(\nu^\star)=P_{\mathrm T}$, which can be efficiently obtained through a one-dimensional bisection search. In either case, the optimal BS precoder is
\begin{equation}\label{eq:w_solution}
    \mathbf w_k^\star=\mathbf w_k(\nu^\star),~\forall k.
\end{equation}

\subsubsection{CT-RIS Phase-Function Update}
For fixed $\mathbf W$ and $\{\mu_k, \lambda_k\}_{k=1}^{K}$, define $q_{k,i}(\mathbf s) \triangleq \eta g_k(\mathbf s) \mathbf f^T(\mathbf s) \mathbf w_i$. The CT-RIS phase-function subproblem is
\begin{subequations}
	\label{eq:phase_subproblem}
	\begin{align}
		\max_{\theta(\cdot)}
		\quad
		&
		F_{\theta}(\theta)
		\label{eq:phase_objective}
		\\[1ex]
		\mathrm{s.t.}\quad
		&
		\theta(\mathbf s)\in[-\pi,\pi),
		\quad
		\forall\,\mathbf s\in\mathcal S,
		\label{eq:phase_feasible_set}
	\end{align}
\end{subequations}
where
\begin{align}
	F_{\theta}(\theta)
	&=
	\sum_{k=1}^{K}
	2\operatorname{Re}
	\left\{
	A_k
	\int_{\mathcal S}
	q_{k,k}(\mathbf s)e^{j\theta(\mathbf s)}
	\,\mathrm d\mathbf s
	\right\}
	\nonumber\\[1ex]
	&\quad
	-
	\sum_{k=1}^{K}
	B_k
	\sum_{i=1}^{K}
	\left|
	\int_{\mathcal S}
	q_{k,i}(\mathbf s)e^{j\theta(\mathbf s)}
	\,\mathrm d\mathbf s
	\right|^2.
	\label{eq:phase_objective_expanded}
\end{align}

To expose its functional structure, let $z(\mathbf s)=e^{-j\theta(\mathbf s)}$. Then, $\mathbf h_k^H \mathbf w_i=\langle z,q_{k,i}\rangle$. Accordingly, \eqref{eq:phase_objective_expanded} can be written as
\begin{align}
	F_{\theta}(z)
	=
	2\operatorname{Re}\{\langle z,a\rangle\}
	-
	\langle z,\mathcal{R}z\rangle,
	\quad
	|z(\mathbf s)|=1,
	\label{eq:functional_form}
\end{align}
where $a(\mathbf s)\triangleq\sum_{k=1}^{K}	A_kq_{k,k}(\mathbf s)$ and $\mathcal{R}z(\mathbf s)\triangleq\sum_{k=1}^{K}	\sum_{i=1}^{K}	B_kq_{k,i}(\mathbf s)	\langle q_{k,i},z\rangle$.
Here, $\mathcal R$ is a self-adjoint positive-semidefinite linear operator, since $\langle z, \mathcal{R}z\rangle = \sum_{k=1}^{K} \sum_{i=1}^{K} B_k |\langle q_{k,i},z\rangle|^2 \geq 0$.

To characterize the functional stationarity condition, consider the variation
$\theta(\mathbf s)\mapsto\theta(\mathbf s)+\epsilon U(\mathbf s)$, where $U(\mathbf s)$ is an arbitrary real-valued function. The CoV gives
\begin{align}
	\left.
	\frac{\mathrm dF_{\theta}}{\mathrm d\epsilon}
	\right|_{\epsilon=0}
	=
	-2
	\int_{\mathcal S}
	U(\mathbf s)
	\operatorname{Im}
	\left\{
	z^*(\mathbf s)
	\left[
	a(\mathbf s) - \mathcal{R}z(\mathbf s)
	\right]
	\right\}
	\,\mathrm d\mathbf s.
\end{align}
Therefore, any stationary phase function satisfies
\begin{align}
	\operatorname{Im}
	\left\{
	z^*(\mathbf s)
	\left[
	a(\mathbf s)-\mathcal{R}z(\mathbf s)
	\right]
	\right\}
	=
	0
	\quad\text{a.e. on }\mathcal S.
	\label{eq:functional_stationarity}
\end{align}
This is an implicit stationarity condition since $\mathcal Rz$ depends on $z$ itself and therefore does not directly yield a monotonic update.

To construct an MM update, choose $\xi \geq \|\mathcal{R}\|_{\mathrm{op}}$ and define
$\mathcal M\triangleq\xi\mathcal I - \mathcal{R} \succeq 0$, where $\mathcal I$ denotes the identity operator. A tight minorizer of $F_{\theta}(z)$ at $z^t$ is
\begin{align}
	G(z|z^t)
	=
	2\operatorname{Re}
	\left\{
	\left\langle
	z,
	a+\mathcal M z^t
	\right\rangle
	\right\}
	+
	C^t,
	\label{eq:MM_surrogate}
\end{align}
where $C^t=-\langle z^t,\mathcal M z^t\rangle-\xi|\mathcal S|$ is independent of $z$. Indeed,
\begin{align}
	F_{\theta}(z)-G(z|z^t)
	=
	\left\langle
	z-z^t,
	\mathcal{M}(z - z^t)
	\right\rangle
	\geq0,
\end{align}
with equality at $z=z^t$.

Maximizing \eqref{eq:MM_surrogate} under
$|z(\mathbf s)|=1$ yields the pointwise update
\begin{align}
	z^{t+1}(\mathbf s)
	=
	\frac{
		\Phi^t(\mathbf s)+\xi z^t(\mathbf s)
	}{
		\left|
		\Phi^t(\mathbf s)+\xi z^t(\mathbf s)
		\right|
	},
	\label{eq:mm_x_update}
\end{align}
where $\Phi^t(\mathbf s) \triangleq a(\mathbf s)-\mathcal{R}z^t(\mathbf s)$. Equivalently, the CT-RIS phase function is updated as
\begin{align}
	\theta^{t+1}(\mathbf s)
	=
	-\arg
	\left(
	\Phi^t(\mathbf s)
	+
	\xi e^{-j\theta^t(\mathbf s)}
	\right).
	\label{eq:mm_phase_update}
\end{align}

The explicit computation of the operator norm can be avoided by choosing
\begin{align}
	\xi
	=
	\sum_{k=1}^{K}
	\sum_{i=1}^{K}
	B_k
	\int_{\mathcal S}
	|q_{k,i}(\mathbf s)|^2
	\,\mathrm d\mathbf s,
	\label{eq:xi_upper_bound}
\end{align}
which satisfies $\xi \geq \|\mathcal{R}\|_{\mathrm{op}}$. Consequently,
\begin{align}
	F_{\theta}(z^{t+1})
	\geq
	G(z^{t+1}|z^t)
	\geq
	G(z^t|z^t)
	=
	F_{\theta}(z^t),
\end{align}
and hence the phase-function update guarantees a nondecreasing transformed objective. The overall AO-CoV-MM procedure is summarized in Algorithm~\ref{alg:ao_cov}.

\begin{algorithm}[t]
\caption{AO-CoV-MM Algorithm for Solving Problem~\eqref{eq:transformed_problem}}
\label{alg:ao_cov}
\begin{algorithmic}[1]
\State \textbf{Input:}
$\mathbf f(\mathbf s)$,
$\{g_k(\mathbf s),\sigma_k^2,\alpha_k\}_{k=1}^{K}$,
$P_{\mathrm T}$, tolerance $\varepsilon_{\rm AO}$, and maximum number of iterations $T_{\max}$.
\State \textbf{Initialize:}
feasible $\{\mathbf w_k^{0}\}_{k=1}^{K}$ and
$\theta^{0}(\mathbf s)$; set $t=0$.
\Repeat
	\State Compute $\{\mathbf h_k^{t}\}_{k=1}^{K}$ using $\theta^{t}(\mathbf s)$.
	\State Update
	$\{\mu_k^{t+1},\lambda_k^{t+1}\}_{k=1}^{K}$
	using \eqref{eq:auxiliary_updates}.
	\State Update $\{\mathbf w_k^{t+1}\}_{k=1}^{K}$
	using \eqref{eq:w_solution}.
	\State Compute
	$\{q_{k,i}(\mathbf s)\}$, $a(\mathbf s)$,
	$\mathcal Rz^t(\mathbf s)$, $\Phi^t(\mathbf s)$, and $\xi$.
	\State Update $\theta^{t+1}(\mathbf s)$ using
	\eqref{eq:mm_phase_update}.
	\State $t\leftarrow t+1$.
\Until{the relative improvement is below $\varepsilon_{\rm AO}$ or $t=T_{\max}$.}
\State \textbf{Return:}
$\{\mathbf w_k^\star=\mathbf w_k^{t}\}_{k=1}^{K}$ and
$\theta^\star(\mathbf s)=\theta^{t}(\mathbf s)$.
\end{algorithmic}
\end{algorithm}

\subsection{Convergence and Computational Complexity}

\subsubsection{Convergence}
According to the update order of the proposed AO-CoV-MM algorithm, the transformed objective values satisfy
\begin{align}
	F_0
	\left(
	\mathbf W^t,\theta^t,
	\boldsymbol{\mu}^t,\boldsymbol{\lambda}^t
	\right)
	&\overset{(a)}{\leq}
	F_0
	\left(
	\mathbf W^t,\theta^t,
	\boldsymbol{\mu}^{t+1},\boldsymbol{\lambda}^{t+1}
	\right)
	\nonumber\\[1ex]
	&\overset{(b)}{\leq}
	F_0
	\left(
	\mathbf W^{t+1},\theta^t,
	\boldsymbol{\mu}^{t+1},\boldsymbol{\lambda}^{t+1}
	\right)
	\nonumber\\[1ex]
	&\overset{(c)}{\leq}
	F_0
	\left(
	\mathbf W^{t+1},\theta^{t+1},
	\boldsymbol{\mu}^{t+1},\boldsymbol{\lambda}^{t+1}
	\right).
	\label{eq:AO_monotonicity}
\end{align}
Here, $(a)$ follows because $\boldsymbol{\mu}^{t+1}$ and $\boldsymbol{\lambda}^{t+1}$ are the optimal auxiliary variables for fixed $\mathbf W^t$ and $\theta^t$; $(b)$ follows from the optimal solution of the BS-precoding subproblem; and $(c)$ follows from the MM-based phase-function update. In particular, the tight minorization property established above guarantees that the phase update does not decrease $F_\theta$, and hence does not decrease $F_0$.

For any fixed $\mathbf W$ and $\theta(\cdot)$,
$F_0$ is no greater than the WSR obtained by maximizing it over the auxiliary variables. Since the transmit power is bounded, the CT-RIS transmission coefficient has unit modulus, and $\sigma_k^2>0$, the WSR is finite. Therefore, the nondecreasing sequence in
\eqref{eq:AO_monotonicity} is upper bounded and converges to a finite value. This establishes convergence of the objective-value sequence, although global optimality is not guaranteed.

\subsubsection{Computational Complexity}
The surface integrals are evaluated using Gauss--Legendre quadrature with $M\times M$ quadrature nodes. Reconstructing the equivalent channels requires $\mathcal O(M^2KN)$ operations, while updating the auxiliary variables requires $\mathcal O(K^2N)$ operations. Constructing $\mathbf Q$ requires $\mathcal O(KN^2)$ operations. Let $T_b$ denote the number of bisection iterations used to determine $\nu^\star$; the BS-precoder update then has complexity $\mathcal O(KN^2+T_bN^3)$. For the phase-function update, the products $\{\mathbf f^T(\mathbf s)\mathbf w_i\}$ can be reused across the users. Consequently, constructing $\{q_{k,i}(\mathbf s)\}$ and evaluating $\mathcal Rz$, $\xi$, and the pointwise phase update require $\mathcal O\!\left(M^2(KN+K^2)\right)$ operations. The additional pointwise phase-alignment operation introduced by MM has complexity $\mathcal O(M^2)$ and therefore does not change the leading complexity order. Hence, after $T_o$ AO iterations, the overall computational complexity is $\mathcal O \left(T_o \left[T_bN^3 + KN^2 + K^2N + M^2(KN+K^2)\right]\right)$.

\section{Numerical Results}
We consider a CT-RIS-enabled system operating at $f_c=2.4$~GHz. The square CT-RIS has a side length of $0.5$~m and is centered at $\mathbf s_0=[0,0.5,10]^T$~m. The BS employs a $4 \times 4$ UPA with half-wavelength antenna spacing centered at $[0,0,0]^T$~m and serves $K=4$ single-antenna users located at $\mathbf r_k=[r_{k,x},r_{k,y},0]^T$~m, where $r_{k,x} \sim \mathcal U[-25,25]$ and $r_{k,y} \sim \mathcal U[20,70]$. Each CT-RIS-to-user link comprises $L_k = 5$ independent scattered paths with $\beta_{k,l} \sim \mathcal{CN}(0,\ell_k^2/L_k)$, where $\ell_k = \lambda_c / (4 \pi d_k)$ denotes the reference large-scale attenuation and $d_k = \|\mathbf r_k - \mathbf s_0\|_2$. The path azimuth and elevation angles are independently generated according to $\phi_{k,l} \sim \mathcal{U}[-\pi/2,\pi/2)$ and $\sin(\vartheta_{k,l}) \sim \mathcal{U}[-1,1]$, respectively \cite{10318061}.

The user weights are set to $\alpha_k=1/K$, making the WSR equal to the average user rate. With a noise power spectral density of $-174$~dBm/Hz and a bandwidth of $20$~MHz, the noise power is $\sigma_k^2 = 7.96\times10^{-14}$~W. The continuous-aperture integrals are evaluated using a $20 \times 20$ Gauss--Legendre quadrature rule \cite{10910020}. The parameter $\xi$ is selected according to \eqref{eq:xi_upper_bound}, and the AO algorithm terminates when the relative objective improvement falls below $\varepsilon_{\rm AO} = 0.001$ or the number of iterations reaches $T_{\max} = 500$. All results are averaged over $N_{\rm MC} = 1000$ independent channel realizations under fixed user locations pre-generated via a random seed, with the same realizations used for all considered schemes.

For performance comparison, we consider the following benchmark schemes:
\begin{itemize}
    \item \textbf{Subaperture CT-RIS:} The physical aperture $\mathcal S$ is partitioned into $N_s$ non-overlapping square subapertures $\{\mathcal S_n\}_{n=1}^{N_s}$ with side length $d_s$, such that $\mathcal S = \bigcup_{n=1}^{N_s}\mathcal S_n$, $|\mathcal S_n| = d_s^2$, and $N_s =\left \lfloor \frac{\,|\mathcal{S}|\,}{ d_s^2}\right \rfloor $. The phase response is constrained to be constant within each subaperture, i.e., $\theta(\mathbf s)=\theta_n$ for $\mathbf s\in\mathcal S_n$, yielding
	\begin{equation}
		\mathbf h_{k,\mathrm{S}}^H
		=
		\eta  \sum_{n=1}^{N_s}
		e^{j\theta_n}
		\int_{\mathcal S_n}
		g_k(\mathbf s)\mathbf f^T(\mathbf s)
		\,\mathrm d\mathbf s.	
	\end{equation}
	Throughout the simulations, we set $d_s = \lambda_c / 2$, such that the subaperture centers form a half-wavelength grid and are used as the element locations in the element-based T-RIS benchmark described below.
    
	\item \textbf{Element-based T-RIS:} 
    Based on the above half-wavelength grid, the conventional element-based T-RIS is constructed by placing one ideal point element at each subaperture center $\mathbf s_n$, with $\theta_n$ denoting the phase shift of the $n$-th element. Its equivalent channel is given by
    \begin{equation}
		\mathbf h_{k,\mathrm{E}}^H
		=
		\eta A_{\mathrm{ref}}
		\sum_{n=1}^{N_s}
		g_k(\mathbf s_n)
		\mathbf f^T(\mathbf s_n)
		e^{j\theta_n}.
	\end{equation}
    Under this setting, the subaperture CT-RIS and the element-based T-RIS have the same physical aperture and the same number of phase control variables. However, their effective receiving areas per phase variable are $d_s^2=\lambda_c^2 / 4$ and $A_{\mathrm{ref}} = \lambda_c^2 / (4 \pi)$, respectively, and hence $d_s^2 = \pi A_{\mathrm{ref}}$. For both benchmarks, the phase variables and BS precoders are jointly optimized following \cite{9570775}.
\end{itemize}

\begin{figure}[t]
    \vskip-5pt
	\centering	\includegraphics[width=0.85\columnwidth]{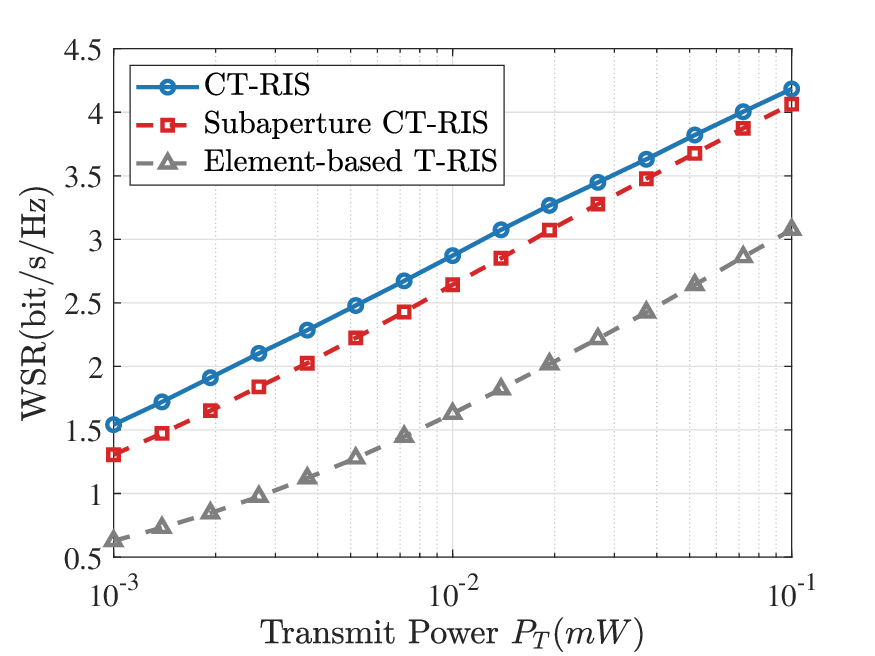}
	\caption{WSR versus transmit power.}
	\label{fig:rate_power}
\end{figure}

Fig.~\ref{fig:rate_power} evaluates the WSR versus transmit power and illustrates the two sources of the performance gain of the proposed CT-RIS over the conventional element-based T-RIS across $P_{\mathrm T} \in [10^{-3}, 10^{-1}]$ mW. The proposed CT-RIS consistently achieves the highest WSR, followed by the subaperture CT-RIS and the element-based T-RIS. The gap between the subaperture CT-RIS and the element-based T-RIS reflects the gain arising from their different effective receiving areas. Meanwhile, the additional gap between the proposed CT-RIS and the subaperture CT-RIS reflects the spatially continuous phase control gain, since the latter constrains the phase response to be constant within each subaperture.

\begin{figure}[t]
    \vskip-5pt
	\centering	\includegraphics[width=0.85\columnwidth]{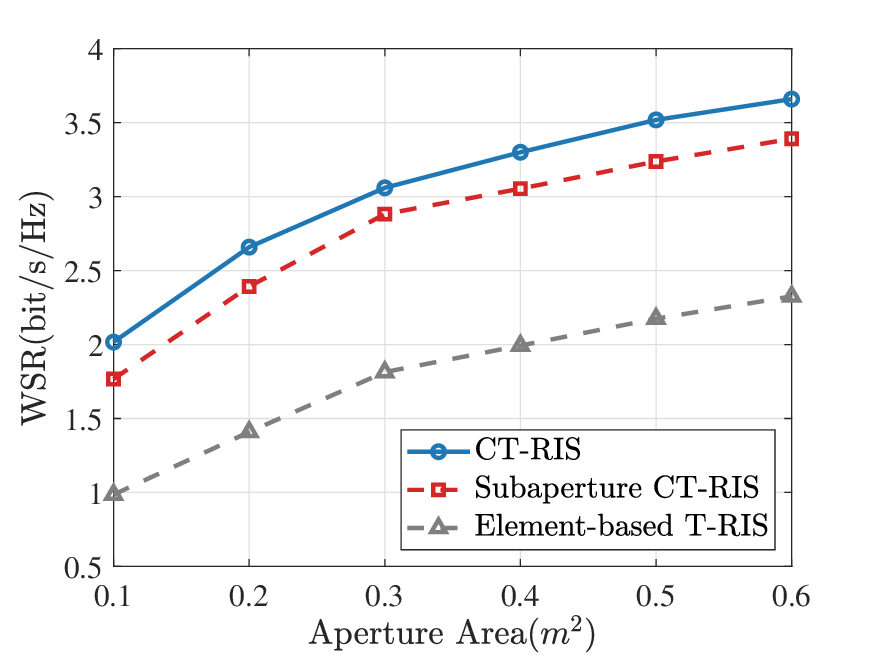}
	\caption{WSR versus aperture size.}
	\label{fig:rate_area}
\end{figure}

The impact of aperture area is illustrated in Fig.~\ref{fig:rate_area}. As the aperture area increases, the WSR of all considered schemes increases monotonically. The proposed CT-RIS consistently achieves the highest WSR, followed by the subaperture CT-RIS and the element-based T-RIS. The performance gaps remain evident across the considered aperture sizes, confirming the contributions of both the effective receiving area and spatially continuous phase control.

\begin{figure}[t]
    \vskip-5pt
	\centering	\includegraphics[width=0.85\columnwidth]{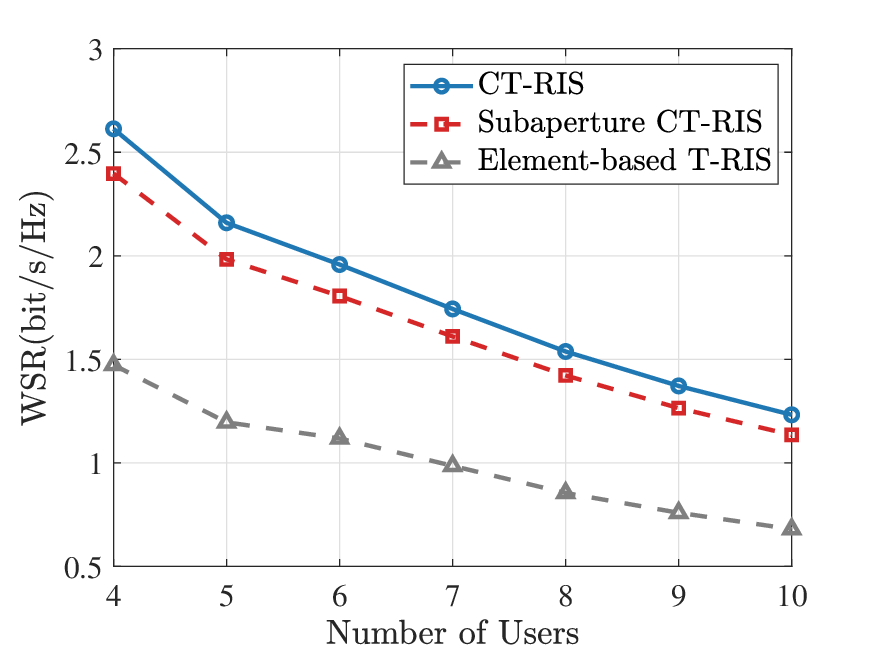}
	\caption{WSR versus the number of users.}
	\label{fig:rate_users}
\end{figure}

Fig.~\ref{fig:rate_users} evaluates the WSR performance versus the number of users $K$. The WSR of all schemes decreases with $K$, reflecting the increased difficulty of multiuser interference management under a fixed total transmit-power budget. Nevertheless, the proposed CT-RIS maintains a clear WSR advantage over both benchmarks across the considered range.

\section{Conclusion}
This paper developed a joint beamforming framework for CT-RIS-enabled multiuser downlink communications, in which the BS precoders and the spatially continuous CT-RIS phase function were jointly optimized to maximize the WSR. The resulting mixed finite- and infinite-dimensional non-convex optimization problem was addressed using an AO-CoV-MM algorithm. A subaperture CT-RIS was introduced as an intermediate benchmark to decompose the performance gain of the proposed CT-RIS over the conventional element-based T-RIS into two sources. Specifically, the proposed CT-RIS outperformed the subaperture CT-RIS, demonstrating the benefit of spatially continuous phase control, while the subaperture CT-RIS outperformed the element-based T-RIS owing to its larger effective receiving area.



\bibliographystyle{IEEEtran}   
\bibliography{bib}

@ARTICLE{9424177,
  author={Liu, Yuanwei and Liu, Xiao and Mu, Xidong and Hou, Tianwei and Xu, Jiaqi and Di Renzo, Marco and Al-Dhahir, Naofal},
  journal={IEEE Communications Surveys \& Tutorials}, 
  title={Reconfigurable Intelligent Surfaces: Principles and Opportunities}, 
  year={2021},
  volume={23},
  number={3},
  pages={1546-1577},
  doi={10.1109/COMST.2021.3077737}}

@ARTICLE{10680462,
  author={Li, Zhendong and Chen, Wen and Wu, Qingqing and Liu, Ziwei and He, Chong and Bai, Xudong and Li, Jun},
  journal={IEEE Vehicular Technology Magazine}, 
  title={Transmissive Reconfigurable Intelligent Surface-Enabled Transceiver Systems: Architecture, Design Issues, and Opportunities}, 
  year={2024},
  volume={19},
  number={4},
  pages={44-53},
  doi={10.1109/MVT.2024.3449794}}

@ARTICLE{10177872,
  author={Tang, Junwen and Cui, Mingyao and Xu, Shenheng and Dai, Linglong and Yang, Fan and Li, Maokun},
  journal={IEEE Transactions on Communications}, 
  title={Transmissive {RIS} for {B5G} Communications: Design, Prototyping, and Experimental Demonstrations}, 
  year={2023},
  volume={71},
  number={11},
  pages={6605-6615},
  doi={10.1109/TCOMM.2023.3292477}}

@ARTICLE{10740042,
  author={Li, Zhendong and Chen, Wen and Zhu, Xusheng and Wu, Qingqing and Ni, Gang and Zhang, Shanshan and Li, Jun},
  journal={IEEE Internet of Things Journal}, 
  title={Transmissive {RIS} Transceiver Enabled Multistream Communication Systems: Design, Optimization, and Analysis}, 
  year={2025},
  volume={12},
  number={5},
  pages={5985-6000},
  doi={10.1109/JIOT.2024.3488799}}

@ARTICLE{11342402,
  author={Guo, Yuan and Chen, Wen and Wu, Qingqing and Li, Zhendong and Wang, Kunlun and Tang, Hongying and Li, Jun},
  journal={IEEE Transactions on Vehicular Technology}, 
  title={Fair Rate Maximization for Multi-User Multi-Cell {MISO} Communication Systems via Novel Transmissive {RIS} Transceiver}, 
  year={2026},
  volume={75},
  number={6},
  pages={11128-11142},
  doi={10.1109/TVT.2026.3651497}}

@ARTICLE{9983541,
  author={Li, Zhendong and Chen, Wen and Zhang, Ziheng and Wu, Qingqing and Cao, Huanqing and Li, Jun},
  journal={IEEE Internet of Things Journal}, 
  title={Robust Sum-Rate Maximization in Transmissive {RMS} Transceiver-Enabled {SWIPT} Networks}, 
  year={2023},
  volume={10},
  number={8},
  pages={7259-7271},
  doi={10.1109/JIOT.2022.3228868}}

@ARTICLE{9136592,
  author={Huang, Chongwen and Hu, Sha and Alexandropoulos, George C. and Zappone, Alessio and Yuen, Chau and Zhang, Rui and Renzo, Marco Di and Debbah, Merouane},
  journal={IEEE Wireless Communications}, 
  title={Holographic MIMO Surfaces for 6G Wireless Networks: Opportunities, Challenges, and Trends}, 
  year={2020},
  volume={27},
  number={5},
  pages={118-125},
  doi={10.1109/MWC.001.1900534}}

@INPROCEEDINGS{9443506,
  author={Pizzo, Andrea and Marzetta, Thomas and Sanguinetti, Luca},
  booktitle={2020 54th Asilomar Conference on Signals, Systems, and Computers}, 
  title={Holographic MIMO Communications Under Spatially-Stationary Scattering}, 
  year={2020},
  volume={},
  number={},
  pages={702-706},
  doi={10.1109/IEEECONF51394.2020.9443506}}

@ARTICLE{8314727,
  author={Shen, Kaiming and Yu, Wei},
  journal={IEEE Transactions on Signal Processing}, 
  title={Fractional Programming for Communication Systems—Part {I}: Power Control and Beamforming}, 
  year={2018},
  volume={66},
  number={10},
  pages={2616-2630},
  doi={10.1109/TSP.2018.2812733}}

@ARTICLE{8310563,
  author={Shen, Kaiming and Yu, Wei},
  journal={IEEE Transactions on Signal Processing}, 
  title={Fractional Programming for Communication Systems—Part {II}: Uplink Scheduling via Matching}, 
  year={2018},
  volume={66},
  number={10},
  pages={2631-2644},
  doi={10.1109/TSP.2018.2812748}}

@ARTICLE{10910020,
  author={Wang, Zhaolin and Ouyang, Chongjun and Liu, Yuanwei},
  journal={IEEE Transactions on Wireless Communications}, 
  title={Beamforming Optimization for Continuous Aperture Array ({CAPA})-Based Communications}, 
  year={2025},
  volume={24},
  number={6},
  pages={5099-5113}
}

@ARTICLE{10318061,
  author={Zhu, Lipeng and Ma, Wenyan and Zhang, Rui},
  journal={IEEE Transactions on Wireless Communications}, 
  title={Modeling and Performance Analysis for Movable Antenna Enabled Wireless Communications}, 
  year={2024},
  volume={23},
  number={6},
  pages={6234-6250},
  doi={10.1109/TWC.2023.3330887}}

@ARTICLE{9570775,
  author={Li, Zhendong and Chen, Wen and Cao, Huanqing},
  journal={IEEE Wireless Communications Letters}, 
  title={Beamforming Design and Power Allocation for Transmissive {RMS}-Based Transmitter Architectures}, 
  year={2022},
  volume={11},
  number={1},
  pages={53-57},
  doi={10.1109/LWC.2021.3120284}}

\end{document}